\documentclass[epj]{svjour}
\usepackage{graphics}
\begin{document}
\title{Analysis of the image of pion-emitting sources in source center of mass frame}
\author{Yanyu Ren\inst{1}
\thanks{\emph{email:ryy@hit.edu.cn} }
\and Qichun Feng\inst{1}
\and Weining Zhang\inst{1,2}
\and Lei Huo\inst{1}
\and Jingbo Zhang\inst{1}
\and Jianli Liu\inst{1}
\and Guixin Tang\inst{1}
%
}                     
\offprints{}          
\institute{Department of Physics, Harbin Institute of Technology,
Harbin, Heilongjiang 150006, China \and School of Physics and Optoelectronic Technology, Dalian University
of Technology, Dalian, Liaoning 116024, China}
\date{Received: date / Revised version: date}
%
\abstract{
In this paper, we try a method to extract the image of pion-emitting
source function in the center-of-mass frame of source (CMFS). We choose the identical pion pairs according to the
difference of their energy and use these pion pairs to build the
correlation function. The purpose is to reduce the effect of
$\triangle E \triangle t$, thus the corresponding imaging result can
tend to the real source function. We examine the effect of this method by comparing
its results with real source functions extracted from models directly.
} 
\maketitle
\section{Introduction}
\label{intro}
Two-pion Hanbury-Brown-Twiss (HBT) interferometry is a valid tool for
probing the space-time structure of the particle-emitting sources in high energy heavy ion collisions
\cite{UAW99,RMW00,MAL05}. For getting the space-time information of the
sources, people have developed many methods to analyze the
interferometry results. These methods can be summarized into two
categories. In conventional HBT analysis one needs to fit the
correlation functions with a parametrized Guassian formula
\cite{UAW99,Cha95,Spr98,UHE99}; while the imaging technique
introduced by Brown and Danielewicz \cite{DAB97,DAB98,DAB01} can
obtain the two-pion source function directly from
the HBT correlation functions.

In conventional HBT analysis, the corresponding HBT results are
model dependent, because people have to introduce a Gaussian emission
function before fitting the HBT parameters. If the particle-emitting
sources produced in relativistic heavy ion collisions are far from
Gaussian distributed
\cite{PHE07,PCH05,PCH07,PCH08,PHE08,RAL08,ZTY09,ZWL02,ZWL04,TCS04,WNZ06,YYR08},
this conventional HBT method is inappropriate
\cite{UHE99,TCS04,SNI98,DHA00,EFR06}. In contrast, the imaging
technique is a model-independent method. It has been
developed and used in analyzing one- and multi-dimensional source
geometry in relativistic heavy ion collisions
\cite{DAB01,PHE07,PCH05,PCH07,PCH08,PHE08,RAL08,ZTY09,SYP01,GVE02,PCH03,PDA04,DAB05,PDA05,PDA07,ZWN09}.

The imaging analysis is performed in the center-of-mass frame of the pion pair (CMFP) at present. Because of the different velocities of different pion pairs, the geometry meaning of source function extracted by the imaging method in the CMFP is cryptic and distorted.  However, the source image extracted in the center-of-mass frame of the source (CMFS) will be affected by the duration of particle emission.  In this paper, we examine the source images extracted in the CMFS.  We implement a cut to the energy difference of the pion pair to decrease the duration effect on the source image.  It is found that the source image extracted in the CMFS with the energy cut $\Delta E <10$ MeV is quite the same as the real source function.

\section{Extracting two-pion source functions in the CMFS}
\begin{figure} [h]
\resizebox{0.4\textwidth}{!}{%
  \includegraphics{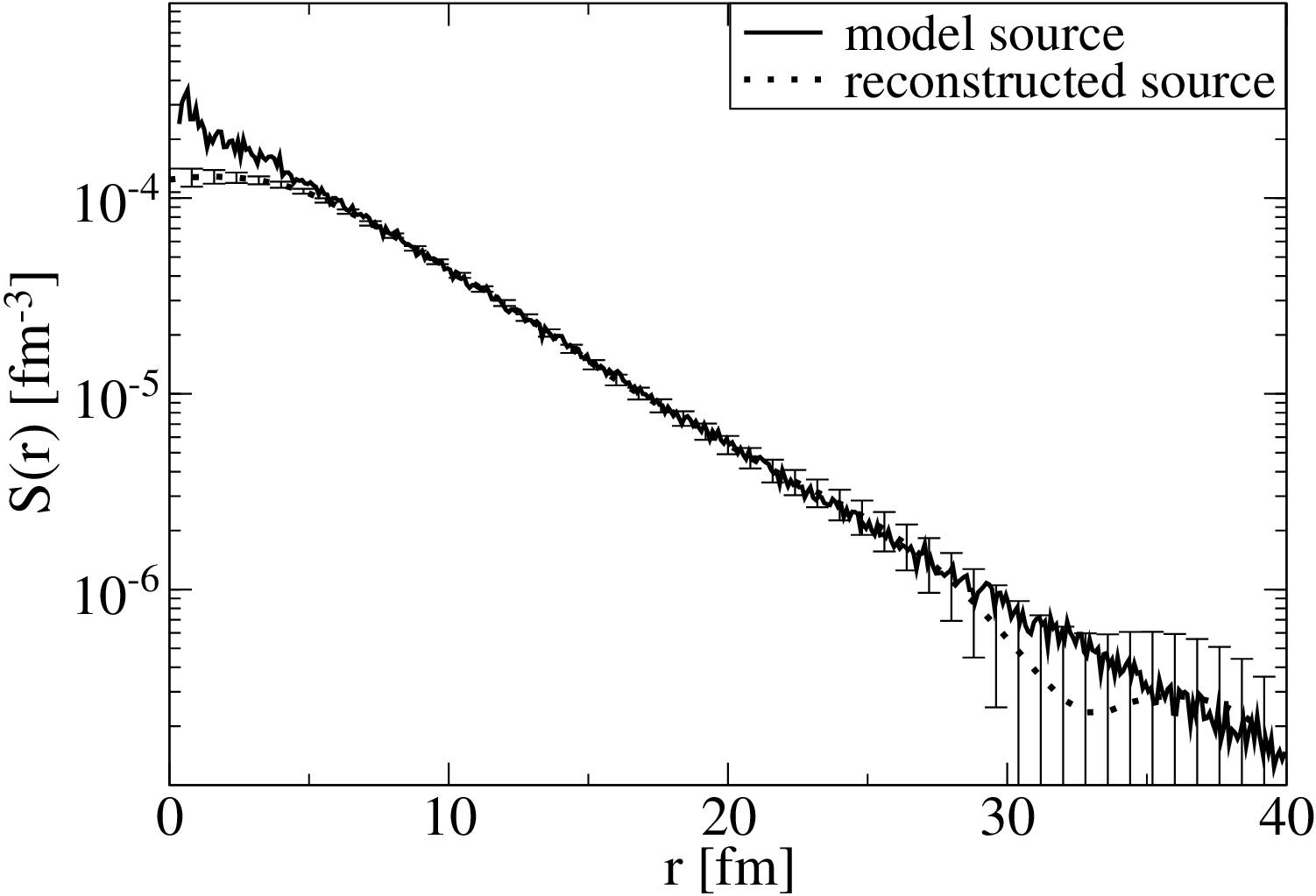}
}
\caption{Comparison between the model source (solid curve) and the reconstructed source (dotted curve) with the AMPT model in the CMFP.} \label{1}
\end{figure}

The imaging technique can be generalized as follows
\cite{DAB97,DAB01}. Ignoring the interactions between the pions, the two-pion correlation function can be
represented as
\cite{UAW99}
\begin{eqnarray}
\label{all bk0} C({\bf q})=1+\int d^4{x}\cos({\bf q}\cdot{\bf
r}-{\bigtriangleup E}{\bigtriangleup t})d(x),
\end{eqnarray}
where $d(x)$ is the so-called relative distance distribution.
 If we write this equation in the CMFP, the value of
${\bigtriangleup E}{\bigtriangleup t}$ is zero. Then, Eq. (\ref{all bk0}) can be written as
\begin{eqnarray}
\label{all bk01} C({\bf q})=1+\int \cos({\bf q}{\cdot}{\bf r})
S({\bf r}) d {\bf r}.
\end{eqnarray}
Here $S({\bf r})=\int d(x) dt$ is the source function which describes the distribution of the relative
separation of emission points for two particles. The
angle-averaged version of Eq. (\ref{all bk01}) is
\begin{eqnarray}
\label{all bk1}\mathcal {R}({q})&=&C({q})-1
=4\pi\int\frac{1}{{q}}{\sin({q}r)}{r}S(r)dr.
\end{eqnarray}
Finally, the one-dimensional source function $S(r)$ can be
calculated by Fourier transform
\begin{eqnarray}
\label{all bk2} S(r)=\frac{1}{2\pi^2}\frac{1}{r}\int\mathcal
{R}({q}){q}\sin({q}{r})d{q}.
\end{eqnarray}
From above equations, we can see that the imaging technique is a
model-independent way to obtain the source function in CMFP.
The way of testing the validity of imaging technique is comparing the reconstructed source with the model source\cite{DAB01,DAB05}. The reconstructed source is the $S(r)$
calculated by imaging technique. The model source is gained by computing pairs of pions generated from the source
with relative momentum less than $60$ MeV/$c$ \cite{DAB05,PAN99}. Now we take the String Melting
AMPT(A Multi-Phase Transport) model\cite{ZWL05} for example, the source function of which has been analyzed by two-pion correlation functions\cite{ZWL09,Shan09}. We simulate Au+Au collisions source for the $\sqrt{s_{_{NN}
}}=200$ GeV with $b=0$ fm, the number of events is 200. From Fig.
\ref{1} we can give corresponding result: the solid curve is computed by all
the $\pi^{+}$ pairs which relative momentum is less than $60$~MeV/$c$ (similarly hereinafter); the dotted curve is calculated from correlation function with imaging technique. We can see that two curves are in good agreement, which means the imaging technique can work well in CMFP. The above is the brief summary of current one-dimensional imaging technique.

\vspace{0mm}
\begin{figure} [h]
\resizebox{0.4\textwidth}{!}{%
  \includegraphics{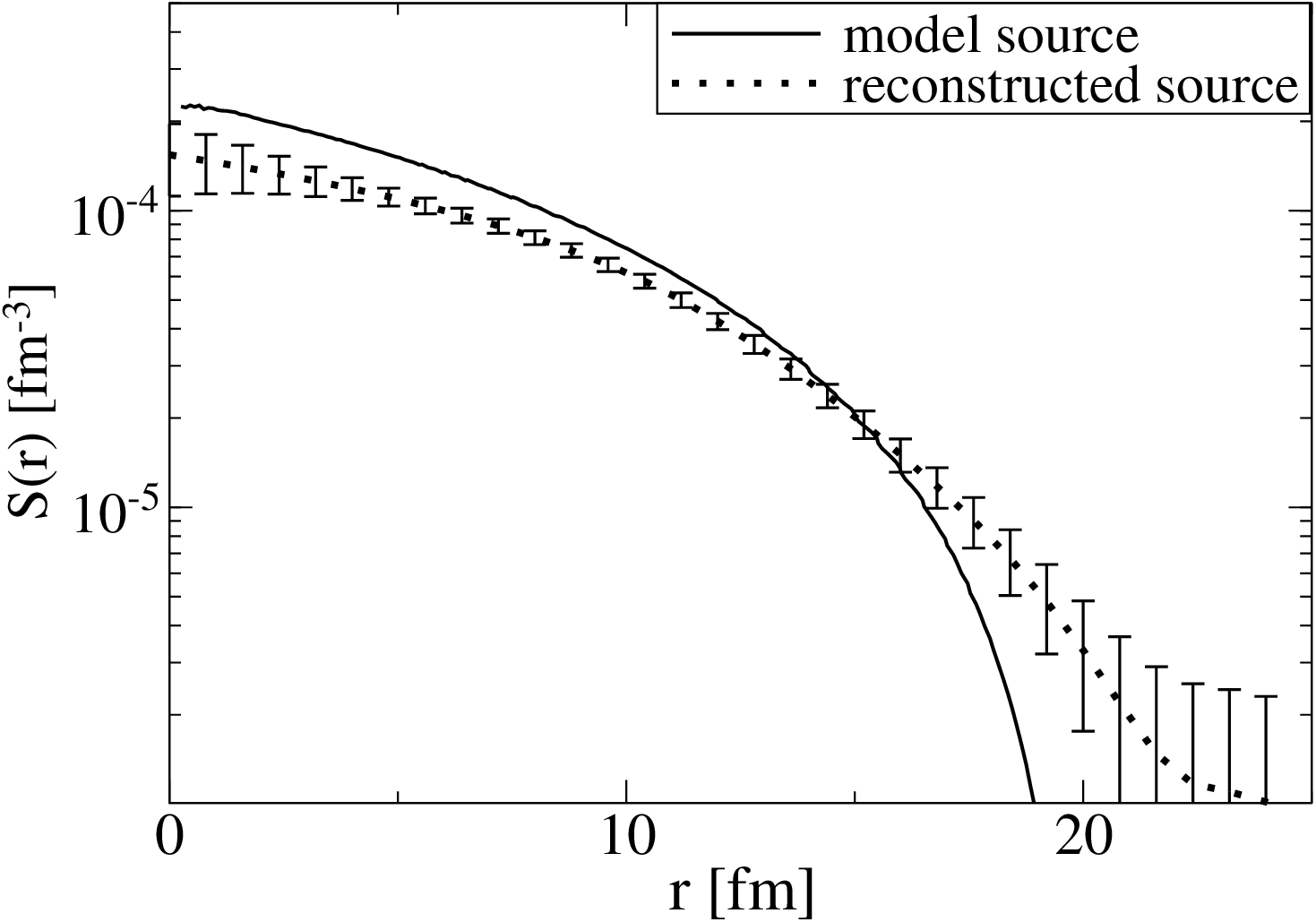}
}
\vspace{3mm}
\caption{Comparison between the model source (solid curve) and the reconstructed source (dotted curve) with a homogeneous spherical model in the CMFS.} \label{2}
\end{figure}

However, the CMFP is not a fixed frame, since different pair of pions has different center of mass velocity.
In this paper, we try to improve this technique and intend to extract the source function $S(r)$ in the CMPS.
To
achieve this point, it is natural for us to use the Eq. (\ref{all bk0})-(\ref{all bk2}) in the CMFS. We take the simple
homogeneous spherical pion-emitting source with radius $r=10$ fm as the preliminary analysis
object and set the momentum
for the Boltzmann distribution ($T_f=158$ MeV). The purpose of setting such a simple source is not to
conform experimental data, but convenient for us to analyze. In
particular, we set the duration of particle emission time also in
Gaussian distribution
\begin{eqnarray}
\label{all bk555}
f(t)=
\left\{
\begin{array}{l}
\sqrt{{2}/{\pi}}\exp\left(-{t^2}/{2\tau^2}\right)/\tau~~~~~~~~~~~t>0,\\
\\
~~~~~~~~~~~~0~~~~~~~~~~~~~~~~~~~~\hspace{3.6mm}~~~~~t<0.
\end{array}\right.
\end{eqnarray}
We set the standard deviation with $\tau=10$ fm/$c$ (similarly hereinafter).
In Fig. \ref{2},
we compare the model source (solid curve) with the reconstructed
source (dotted curve). Here the model source is gained by computing $10^7$ pairs of pions generated from the source.
It can be seen that there
exists obvious deviation. The reason is obvious: although the Eq. (\ref{all bk0}) can be written in any frame, $\triangle E \triangle
t$ is not zero in the CMFS, which make the Eq. (\ref{all bk01}) inaccurate. To solve this problem, we intend to cut the pion pairs according to the
difference of their energy and use the remaining pion pairs to build the correlation functions in CMFS. The reasons are as follows: For one
thing, the Eq.
(\ref{all bk01}) is close to accurate when the difference of energy tends to zero. For another, the energy of
particle can be observed in laboratory. So this method can be used to analyse the experimental data.

\vspace{4mm}
\begin{figure} [h]
\vspace{0mm}
\resizebox{0.35\textwidth}{!}{%
  \includegraphics{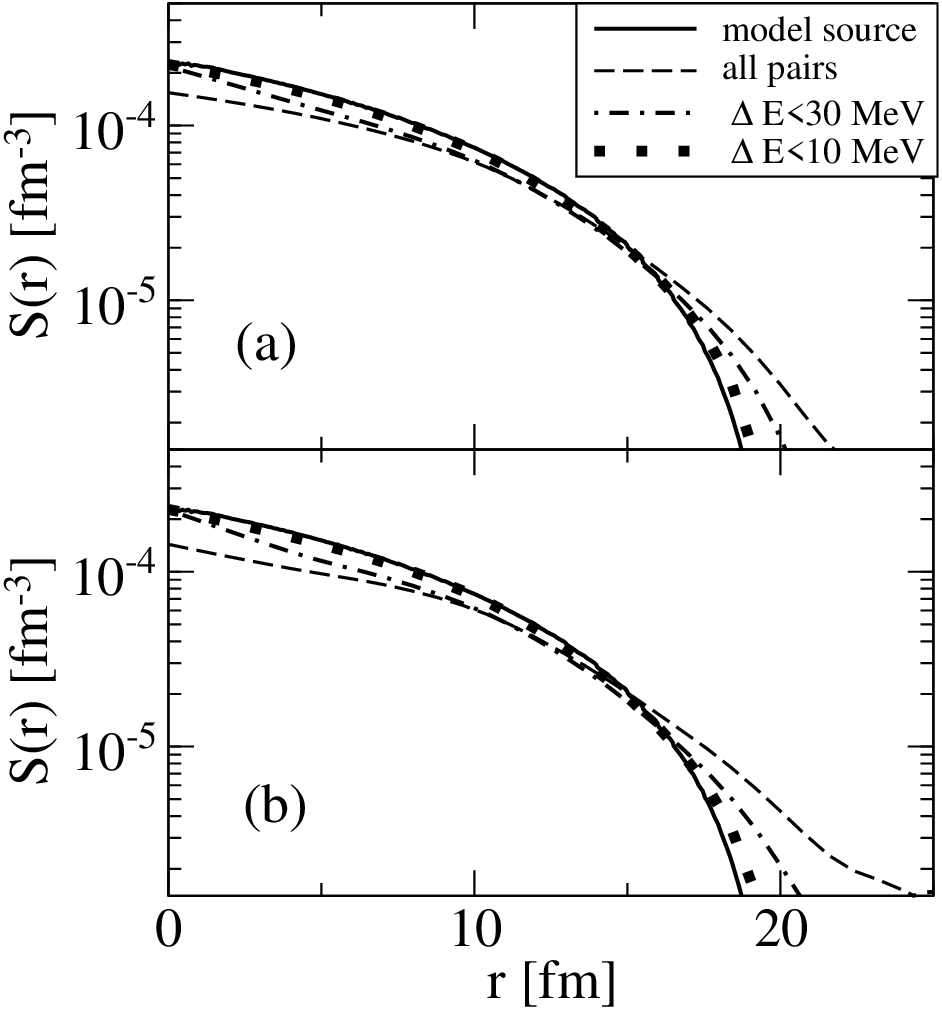}
}
\vspace{6mm}
\caption{Correction of imaging results in the CMFS by cutting
pairs according to their energy difference. In Fig. \ref{3}(a) we
use the same model as in Fig. \ref{2}, while in Fig.
\ref{3}(b) the momentum distribution is changed to a uniform form
with $0$ to $1$ GeV/$c$. } \label{3}
\end{figure}

In Fig. \ref{3}, we plot the reconstructed sources in the CMFS with the cuts of the energy difference of pion pair ($\triangle E$) less than $30$ MeV and $10$ MeV. (The dashed curve, dash-dotted curve and dotted curve
are all imaging results. Due to
the large number of curves, their error bars are not given in this
figure.) From this figure, it can be seen that the difference between
the imaging result and the real model source becomes smaller and
smaller as the $\triangle E$ reduce. And we also find that when
$\triangle E$ is set within $10$ MeV, this difference can almost be
ignored, because $\triangle E \triangle t$ is very small in this
situation. In Fig. \ref{3} (b) we set the momentum to uniform
distribution and get the corresponding result. We can see this
method of $\triangle E$ cut is little influenced by momentum
spectrum.

\vspace{5mm}
\begin{figure} [h]
\resizebox{0.35\textwidth}{!}{%
  \includegraphics{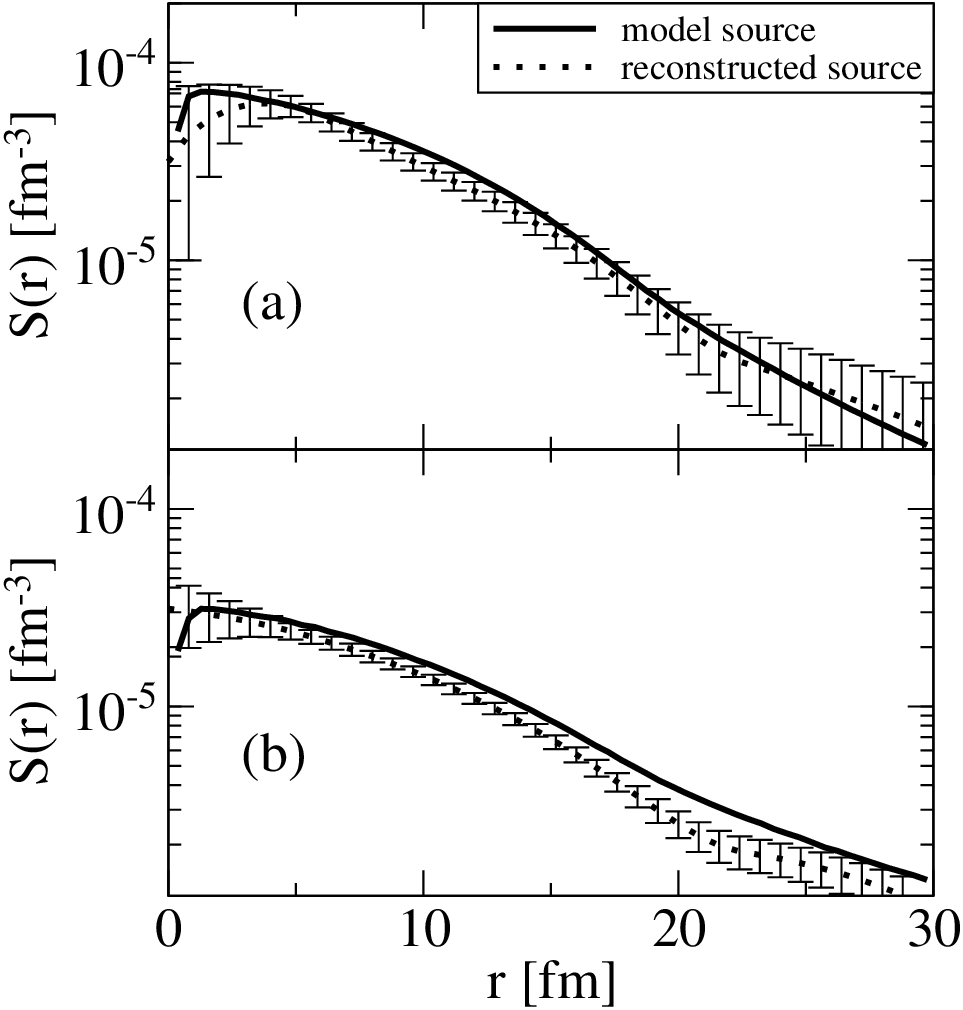}
}
\vspace{7mm}
\caption{The same result of Fig. \ref{3} but in the CMFP.
The momentum is set for Boltzmann distribution (a) and uniform
distribution from $0$ to $1$ GeV/$c$ (b). } \label{4}
\end{figure}

For further discussing the feature of source function in the CMFS, in Fig. \ref{4} we give the corresponding results of Fig.
\ref{3} but in the CMFP. Comparing with these results, we find
two features claim out attention. First, in Fig. \ref{3}(a) and
(b), we can see that the model source functions (solid curves) are
almost exactly the same. While in Fig. \ref{4}(a) and (b), the
model source functions (solid curves) exist obvious difference with
different distribution of momentum. We think the reason is that they
have the same source functions in a fixed frame. But
this source function must be changed in the CMFP for the
Lorentz transform. Obviously, the velocity of every pion pair must
be influenced by the distribution of momentum. And then the source
function will be influenced by the distribution of momentum in the
CMFP. But the source function in the CMPS
only depend on the scale of the particle emitting source and is
little influenced by the distribution of momentum. Second, in
the CMFS there does not exist the pairs of particles which
distance is more than the diameter of the source obviously. This
character can be seen in Fig. \ref{3}(a) and (b) where the radius
is set to $10$ fm. But in Fig. \ref{4}(a) and (b) the source
functions does not reduce to zero at the distance even more than
$25$ fm. This means the source function in the CMFS can
reflect the scale of the particle emitting source more directly.

\section{Testing the imaging results in the CMFS}

\vspace{0mm}
\begin{figure} [h]
\resizebox{0.4\textwidth}{!}{%
  \includegraphics{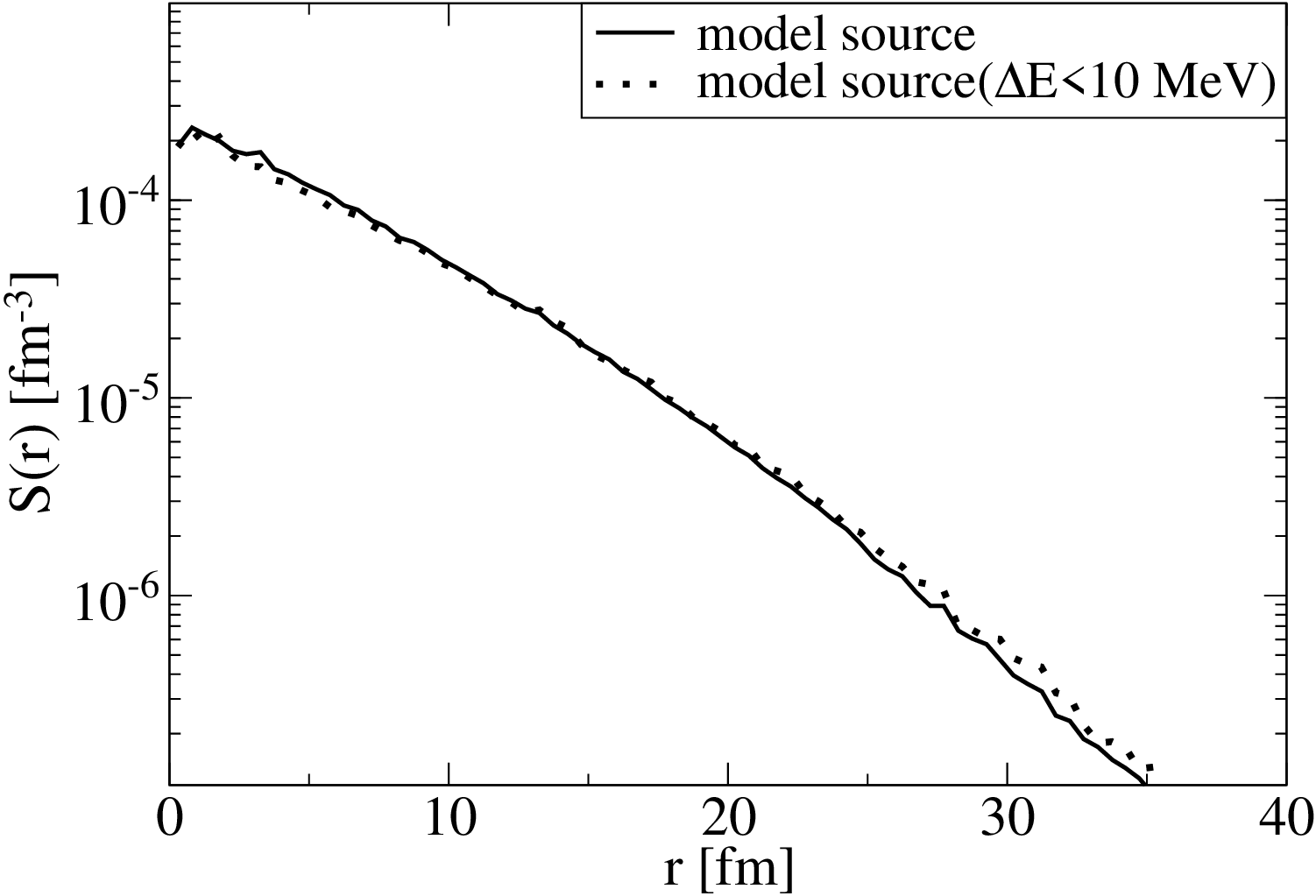}
}
\caption{ The model source function $S(r)$ obtained from the
AMPT model with $b=0$ fm in the CMFS.} \label{5}
\end{figure}

\begin{figure} [h]
\resizebox{0.35\textwidth}{!}{%
  \includegraphics{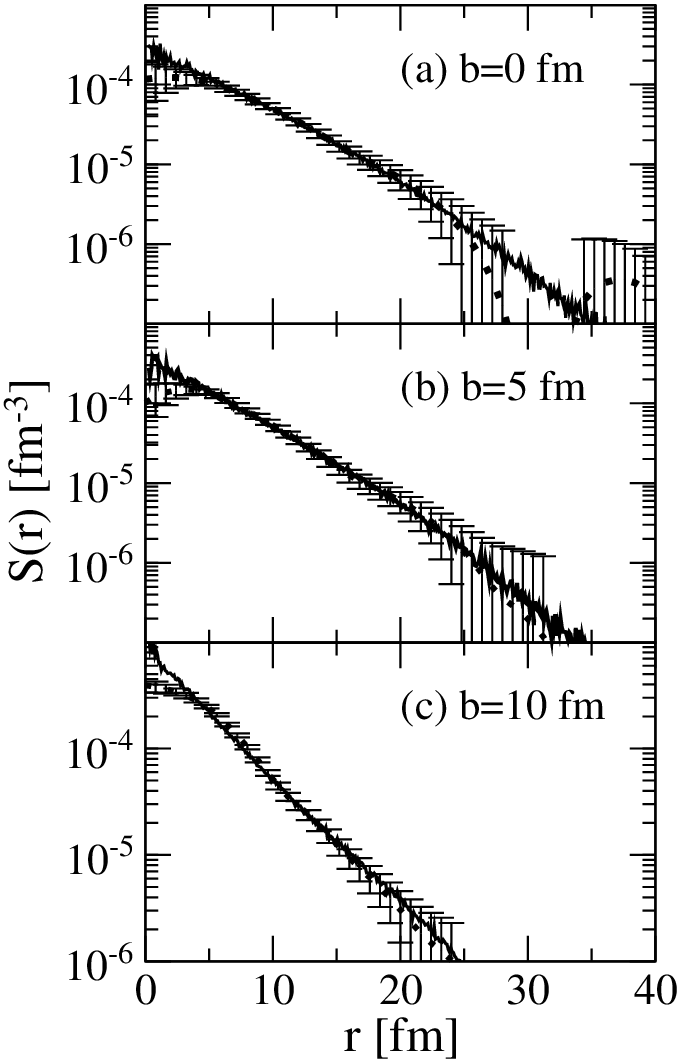}
}
\vspace{4mm}
\caption{ The comparison between the model source (solid curve) and the reconstructed source (dotted curve) in the CMFS. We set impact
parameter (a) $b=0$ fm, (b) $b=5$ fm and (c) $b=10$ fm.} \label{6}
\end{figure}

However, the problem is not the end. In our approach, the $\triangle t$ is gotten by counting the freeze out time in the CMFS. Of course, the $\triangle t$ is not zero and cannot be counted by experimental data. For removing the influence of $\triangle t$, we have to use the pion pairs with very small $\triangle E$ to build the correlation functions. More specifically, our investigation can only reconstruct the $S(r)$ of these pairs. Although this cut of pion pairs is according to the energy, it will influence
the distribution of $S(r)$ more or less. Therefore, we must check
the effect of this scheme with more realistic event
generators. In Fig.
\ref{5} we give the model source functions with AMPT model. The situation is just the same as Fig.
\ref{1} but in the CMFS. The solid curve is the true model source. The dotted curve is also the model source which only computed by
the $\pi^{+}$ pairs with $\bigtriangleup
E<10$~MeV. It can be seen that the influence of $\bigtriangleup
E$ cut on $S(r)$ is very small. In Fig. \ref{6}, we compare the model source and the reconstructed source (also with $\bigtriangleup
E<10$~MeV) with different impact parameters. The number of events is 200 ($b=0$ fm),
500 ($b=5$ fm) and 10000 ($b=10$ fm), respectively. It can be seen that
the two curves are in good agreement, except for the disagreement between the two curves in Fig.6(a) around $r=30$ fm.
The actually obtained correlation function is discrete but not continuous, which is the reason for this disagreement\cite{DAB01}.
We also tried to get the reconstructed source with smaller $\bigtriangleup
E$, the result of which is the same as Fig. \ref{6}. This means $\bigtriangleup
E<10$~MeV is small enough for giving the correct imaging results.

\section{Summary}

In this paper, we try to improve the imaging method for reconstructing the source
function in the CMFS and obtain the following conclusions:
Firstly, the source function in the CMFS can
reflect the scale of the particle emitting source more directly than the CMFP.
Secondly, we give a method to
obtain the source function in the CMFS by cutting the pairs
according to the difference of their energy which can be observed in laboratory. Using various models to
test this method, we find when the energy difference $\bigtriangleup
E<10$~MeV, the reconstructed source is a good approximation to the true model
source.
Finally, it is still a preliminary method which ignores
interactions between the pions and is limited to one-dimensional problem
at present. It means this method can only deal with the correlation function which is constructed by the Coulomb corrected experimental data.
Further investigation of analysis and verification the
possibility of getting more abundant and reliable information in the
CMFS will also be of interest.

\end{document}